# Exact Free Distance and Trapping Set Growth Rates for LDPC Convolutional Codes

David G. M. Mitchell*, Ali E. Pusane†, Michael Lentmaier‡, and Daniel J. Costello, Jr.*
*Dept. of Electrical Engineering, University of Notre Dame, Notre Dame, Indiana, USA,
{david.mitchell, costello.2}@nd.edu
†Dept. of Electrical and Electronics Engineering, Bogazici University, Istanbul, Turkey,
ali.pusane@boun.edu.tr
‡Vodafone Chair Mobile Communications Systems, Dresden University of Technology, Dresden, Germany,
michael.lentmaier@ifn.et.tu-dresden.de

***Abstract*—Ensembles of $(J, K)$-regular low-density parity-check convolutional (LDPCC) codes are known to be *asymptotically good*, in the sense that the minimum free distance grows linearly with the constraint length. In this paper, we use a protograph-based analysis of terminated LDPCC codes to obtain an upper bound on the free distance growth rate of ensembles of periodically time-varying LDPCC codes. This bound is compared to a lower bound and evaluated numerically. It is found that, for a sufficiently large period, the bounds coincide. This approach is then extended to obtain bounds on the trapping set numbers, which define the size of the smallest, non-empty trapping sets, for these asymptotically good, periodically time-varying LDPCC code ensembles.**

## I. Introduction

Low-density parity-check convolutional (LDPCC) codes [1] have been shown to be capable of achieving capacity-approaching performance with iterative message-passing decoding [2]. The excellent iterative decoding thresholds [3], [4] that these codes display has recently been attributed to the *threshold saturation* effect [5]. In addition to good threshold performance, it can also be shown that the minimum free distance typical of most members of these LDPCC code ensembles grows linearly with the constraint length as the constraint length tends to infinity, i.e., they are *asymptotically good* [6]. A large free distance growth rate indicates that codes randomly drawn from the ensemble should have a low error floor under maximum likelihood (ML) decoding.

When sub-optimal decoding methods are employed, there are other factors that affect the performance of a code. For example, it has been shown that so-called 'trapping sets' are a significant factor affecting decoding failures of LDPC codes over the AWGN channel with iterative message-passing decoding. Trapping sets, graphical sub-structures existing in the Tanner graph of LDPC codes, were first studied in [7]. Known initially as *near-codewords*, they were used to analyse the performance of LDPC codes in the error floor, or high signal-to-noise ratio (SNR) region, of the bit error rate (BER) curve. In [8], Richardson developed these concepts and proposed a two-stage technique to predict the error floor performance of LDPC codes based on trapping sets.

In this paper, we use a protograph-based analysis of terminated LDPCC codes to form an upper bound on the free distance growth rate of ensembles of periodically time-varying LDPCC codes. The free distance growth rate can also be bounded below by using ensembles of tail-biting LDPCC codes [9], [10]. By comparing and evaluating these bounds we find that, for a sufficiently large period, the bounds coincide, giving us exact values for the convolutional code free distance growth rates. This approach is then extended to obtain bounds on the trapping set numbers, which define the size of the smallest, non-empty trapping sets, for these asymptotically good, periodically time-varying LDPCC code ensembles. We also show that the trapping set numbers grow linearly with constraint length. For all the ensembles considered, we find that the distance and trapping set growth rates exceed those of corresponding block code ensembles.

This work was partially supported by NSF Grant CCF08-30650.

## II. Background

A protograph [11] is a small bipartite graph that is used to derive a larger graph by taking an $N$-fold graph cover [12], or "lifting", of the protograph. It is an important feature of this construction that each lifted code inherits the degree distribution and graph neigbourhood structure of the protograph. The protograph can be represented by a *base* biadjacency matrix $\mathbf{B}$, where $B_{x,y}$ is taken to be the number of edges connecting variable node $v_y$ to check node $c_x$. The parity-check matrix $\mathbf{H}$ of a protograph-based LDPC block code can be created by replacing each non-zero entry in $\mathbf{B}$ by a sum of $B_{x,y}$ permutation matrices of size $N \times N$ and each zero entry by the $N \times N$ all-zero matrix. The ensemble of protograph-based LDPC block codes with block length $n = Nn_v$ is defined by the set of matrices $\mathbf{H}$ that can be derived from a given protograph by choosing all possible combinations of $N \times N$ permutation matrices.

### A. Convolutional protographs

An ensemble of unterminated LDPCC codes can be described by a *convolutional protograph* [4] with base matrix

$$\mathbf{B}_{[0,\infty]} = \begin{bmatrix} \mathbf{B}_0 & & \\ \mathbf{B}_1 & \mathbf{B}_0 & \\ \vdots & \mathbf{B}_1 & \ddots \\ \mathbf{B}_{m_s} & \vdots & \ddots \\ & \mathbf{B}_{m_s} & \\ & & \ddots \end{bmatrix}, \quad (1)$$

where $m_s$ denotes the syndrome former memory of the convolutional codes and the $b_c \times b_v$ *component base matrices* $\mathbf{B}_i$, $i = 0, \ldots, m_s$, represent the edge connections from the $b_v$ variable nodes at time $t$ to the $b_c$ check nodes at time $t + i$. An ensemble of (in general) time-varying LDPCC codes can then be formed from $\mathbf{B}_{[0,\infty]}$ using the protograph construction method described above, resulting in the associated parity-check matrix

$$\mathbf{H}_{[0,\infty]} = \left[\begin{array}{cccccc} \mathbf{H}_0(0) & & & & & \\ \mathbf{H}_1(1) & \mathbf{H}_0(1) & & & & \\ \vdots & \vdots & & \ddots & & \\ \mathbf{H}_{m_s}(m_s) & \mathbf{H}_{m_s-1}(m_s) & \cdots & \mathbf{H}_0(m_s) & & \\ & \mathbf{H}_{m_s}(m_s+1) & \mathbf{H}_{m_s-1}(m_s+1) & \cdots & \mathbf{H}_0(m_s+1) & \\ & \ddots & \ddots & & \ddots & \end{array}\right].$$

A rate $R = 1 - Nb_c/Nb_v = 1 - b_c/b_v$ time-varying LDPCC code with parity-check matrix $\mathbf{H}_{[0,\infty]}$ is periodically time-varying with period $T$ if $\mathbf{H}_i(t)$ is periodic, i.e., $\mathbf{H}_i(t) = \mathbf{H}_i(t+T), \forall\, i, t$, and if $\mathbf{H}_i(t) = \mathbf{H}_i, \forall\, i, t$, the code is *time-invariant*. We call $\nu_s = N(m_s+1)b_v$ the *decoding constraint length*.

Starting from the base matrix $\mathbf{B}$ of a block code ensemble, one can construct LDPCC code ensembles with the same computation trees. This is achieved by an *edge spreading* procedure (see [4] for details) that divides the edges from each variable node in the base matrix $\mathbf{B}$ among $m_s+1$ component base matrices $\mathbf{B}_i$, $i = 0, \ldots, m_s$, such that the condition $\mathbf{B}_0 + \mathbf{B}_1 + \cdots + \mathbf{B}_{m_s} = \mathbf{B}$ is satisfied. For example, a (3,6)-regular LDPCC ensemble with $m_s = 2$ can be formed from the block base matrix $\mathbf{B} = [\,3 \;\; 3\,]$ by defining the component base matrices $\mathbf{B}_0 = [\,1 \;\; 1\,] = \mathbf{B}_1 = \mathbf{B}_2$.

## III. Termination of LDPCC codes

Suppose that we start the convolutional code with parity-check matrix defined in (1) at time $t = 0$ and terminate it after $L$ time instants. The resulting finite-length base matrix is then given by

$$\mathbf{B}_{[0,L-1]} = \left[\begin{array}{ccc} \mathbf{B}_0 & & \\ \vdots & \ddots & \\ \mathbf{B}_{m_s} & & \mathbf{B}_0 \\ & \ddots & \vdots \\ & & \mathbf{B}_{m_s} \end{array}\right]_{(L+m_s)b_c \times Lb_v} . \quad (2)$$

The matrix $\mathbf{B}_{[0,L-1]}$ can be considered as the base matrix of a terminated protograph-based LDPCC code ensemble. Termination in this fashion results in a rate loss. The design rate of the terminated code ensemble is given as

$$R_L = 1 - \left(\frac{L+m_s}{L}\right)\frac{b_c}{b_v} = 1 - \left(\frac{L+m_s}{L}\right)(1-R)\,, \quad (3)$$

where $R = 1 - Nb_c/Nb_v = 1 - b_c/b_v$ is the rate of the unterminated convolutional code ensemble. Note that, as the *termination factor* $L$ increases, the rate increases and approaches the rate of the unterminated convolutional code ensemble.

The convolutional base matrix $\mathbf{B}_{[0,\infty]}$ can also be terminated using *tail-biting* [13], [14]. Here, for any $\lambda \geq m_s$, the last $b_c m_s$ rows of the terminated parity-check matrix $\mathbf{B}_{[0,\lambda-1]}$ are removed and added to the first $b_c m_s$ rows to form the $\lambda b_c \times \lambda b_v$ tail-biting parity-check matrix $\mathbf{B}_{tb}^{(\lambda)}$ with tail-biting termination factor $\lambda$. Terminating $\mathbf{B}_{[0,\infty]}$ in such a way preserves the design rate of the ensemble, i.e., $R_\lambda = 1 - \lambda b_c/\lambda b_v = 1 - b_c/b_v = R$, and we see that $\mathbf{B}_{tb}^{(\lambda)}$ has exactly the same degree distribution as the original block base matrix $\mathbf{B}$.

## IV. Free distance analysis of protograph-based LDPCC codes

From a convolutional protograph with base matrix $\mathbf{B}_{[0,\infty]}$, we can form a periodically time-varying $N$-fold graph cover with period $T$ by choosing, for the $b_c \times b_v$ submatrices $\mathbf{B}_0, \mathbf{B}_1, \ldots, \mathbf{B}_{m_s}$ in the first $T$ columns of $\mathbf{B}_{[0,\infty]}$, a set of $N \times N$ permutation matrices randomly and independently to form $Nb_c \times Nb_v$ submatrices $\mathbf{H}_0(t), \mathbf{H}_1(t+1), \ldots, \mathbf{H}_{m_s}(t+m_s)$, respectively, for $t = 0, 1, \ldots, T-1$. These submatrices are then repeated periodically (indefinitely) to form $\mathbf{H}_{[0,\infty]}$ such that $\mathbf{H}_i(t+T) = \mathbf{H}_i(t)$, $\forall i, t$. An ensemble of periodically time-varying LDPCC codes with period $T$, rate $R = 1 - Nb_c/Nb_v = 1 - b_c/b_v$, and decoding constraint length $\nu_s = N(m_s+1)b_v$ can then be derived by letting the permutation matrices used to form $\mathbf{H}_0(t), \mathbf{H}_1(t+1), \ldots, \mathbf{H}_{m_s}(t+m_s)$, for $t = 0, 1, \ldots, T-1$, vary over the $N!$ choices of permutation matrix.

### A. Free distance bounds for LDPCC code ensembles

Consider an ensemble of periodically time-varying LDPCC codes with rate $R = 1 - b_c/b_v$ and period $T$ constructed from a convolutional protograph with base matrix $\mathbf{B}_{[0,\infty]}$ as described above. It is known that the average free distance of this ensemble can be bounded below by the average minimum distance of an ensemble of tail-biting LDPCC codes derived from the base matrix $\mathbf{B}_{tb}^{(\lambda)}$ with termination factor $\lambda = T$ [9], [10]. Here, we show that the average free distance of the convolutional ensemble can also be bounded above by the average minimum distance of the ensemble of terminated protograph-based LDPCC codes derived from the base matrix $\mathbf{B}_{[0,L-1]}$ with termination factor $L = T$.

*Theorem 1:* Consider a rate $R = 1 - b_c/b_v$ unterminated, periodically time-varying LDPCC code ensemble with memory $m_s$, decoding constraint length $\nu_s = N(m_s+1)b_v$, and period $T$ derived from $\mathbf{B}_{[0,\infty]}$. Let $\overline{d}_{min}^{(L)}$ be the average minimum distance of the terminated convolutional code ensemble with block length $n = LNb_v$ and termination factor $L$. Then the ensemble average free distance $\overline{d}_{free}^{(T)}$ of the unterminated convolutional code ensemble is bounded above by $\overline{d}_{min}^{(L)}$ for termination factor $L = T$, i.e.,

$$\overline{d}_{free}^{(T)} \leq \overline{d}_{min}^{(T)}. \quad (4)$$

*Sketch of proof.* There is a one-to-one relationship between members of the periodically time-varying LDPCC code ensemble and members of the corresponding terminated LDPCC code ensemble with termination factor $L = T$. For any such pair of codes, every codeword $\mathbf{x} = [\, x_0 \;\; x_1 \;\; \cdots \;\; x_{LNb_v-1} \,]$ in the terminated code can immediately be seen as a codeword $\mathbf{x}_{[0,\infty]} = [\, x_0 \;\; x_1 \;\; \cdots \;\; x_{LNb_v-1} \;\; 0 \;\; \cdots \,]$ in the unterminated code. It follows that the free distance $d_{free}^{(T)}$ of the unterminated code can not be larger than the minimum distance $d_{min}^{(T)}$ of the terminated code. The ensemble average result $\overline{d}_{free}^{(T)} \leq \overline{d}_{min}^{(T)}$ then follows directly. □

Since there is no danger of ambiguity, we will henceforth drop the overline notation when discussing ensemble average distances.

### B. Free distance growth rates of LDPCC code ensembles

In [15], Divsalar presented a technique to calculate the average weight enumerator for protograph-based block code ensembles. This weight enumerator can be used to test if the ensemble is *asymptotically good*, i.e., if the minimum distance typical of most members of the ensemble is at least as large as $\delta_{min} n$, where $\delta_{min}$ its the *minimum distance growth rate* of the ensemble and $n$ is the block length.

For LDPC convolutional codes, conventionally defined as the null space of a sparse parity-check matrix $\mathbf{H}_{[0,\infty]}$, it is natural to define the free distance growth rate with respect to the decoding constraint length $\nu_s$, i.e., as the ratio of the free distance $d_{free}$ to the decoding constraint length $\nu_s$.[1] By bounding $d_{free}^{(T)}$ using (4), we obtain an upper bound on the free distance growth rate as

$$\delta_{free}^{(T)} = \frac{d_{free}^{(T)}}{\nu_s} \leq \frac{\hat{\delta}_{min}^{(T)} T}{(m_s+1)}, \qquad (5)$$

where $\hat{\delta}_{min}^{(T)} = d_{min}^{(T)}/n = d_{min}^{(T)}/(NTb_v)$ is the minimum distance growth rate of the terminated LDPCC code ensemble with termination factor $L = T$ and base matrix $\mathbf{B}_{[0,T-1]}$.[2] Similarly, it was shown in [9] that

$$\delta_{free}^{(T)} \geq \frac{\check{\delta}_{min}^{(T)} T}{(m_s+1)}, \qquad (6)$$

where $\check{\delta}_{min}^{(T)}$ is the minimum distance growth rate of the tail-biting LDPCC code ensemble with tail-biting termination factor $\lambda = T$ and base matrix $\mathbf{B}_{tb}^{(\lambda)}$.

### C. Numerical results

As an example, we consider the $(3,6)$-regular LDPCC code ensemble defined in Section II-A. Since the unterminated convolutional code has rate $R = 1/2$, we calculate the upper bound on the free distance of the periodically time-varying LDPCC code ensemble as $\delta_{free}^{(T)} \leq \hat{\delta}_{min}^{(T)} T/3$ using (5) for termination factors $L = T \geq 3$. Figure 1 displays the minimum distance growth rates $\hat{\delta}_{min}^{(L)}$ of the terminated ensembles defined by $\mathbf{B}_{[0,L-1]}$ for $L = 3, 4, \ldots, 21$ that were calculated using the technique proposed in [15] and the associated upper bounds on the convolutional growth rate $\delta_{free}^{(T)} \leq \hat{\delta}_{min}^{(T)} T/3$ for $L = T$. Also shown are the minimum distance growth rates $\check{\delta}_{min}^{(\lambda)}$ of the tail-biting ensembles defined by base matrix $\mathbf{B}_{tb}^{(\lambda)}$ for $\lambda = 3, 4, \ldots, 21$ and the associated lower bounds on the convolutional growth rate $\delta_{free}^{(T)} \geq \check{\delta}_{min}^{(T)} T/3$ for $\lambda = T$ calculated using (6).

We observe that the calculated ensemble tail-biting convolutional code minimum distance growth rates $\check{\delta}_{min}^{(\lambda)}$ remain constant for $\lambda = 3, \ldots, 11$ and then start to decrease as the termination factor $\lambda$ grows, tending to zero as $\lambda$ tends to infinity. Correspondingly, as $\lambda$ exceeds 11, the lower bound calculated for $\delta_{free}^{(T)}$ levels off at $\delta_{free}^{(T)} \geq 0.086$. The calculated terminated convolutional code minimum distance growth rates $\hat{\delta}_{min}^{(L)}$ are large for small values of $L$ (where the rate loss is larger) and decrease monotonically to zero as $L \to \infty$. Using (5) to obtain an upper bound on the free distance growth rate we observe that, for $T \geq 12$, the upper and lower bounds on $\delta_{free}^{(T)}$ coincide, indicating that, for these values of the period $T$, $\delta_{free}^{(T)} = 0.086$, significantly larger than the $(3,6)$-regular LDPC block code minimum distance growth rate $\delta_{min} = 0.023$. In addition, we note that at the point where the bounds coincide, the growth rates for both termination methods also coincide. Recall that the tail-biting ensembles all have rate $1/2$, wheras the rate of the terminated ensembles is a function of the termination factor $L$ given by (3).

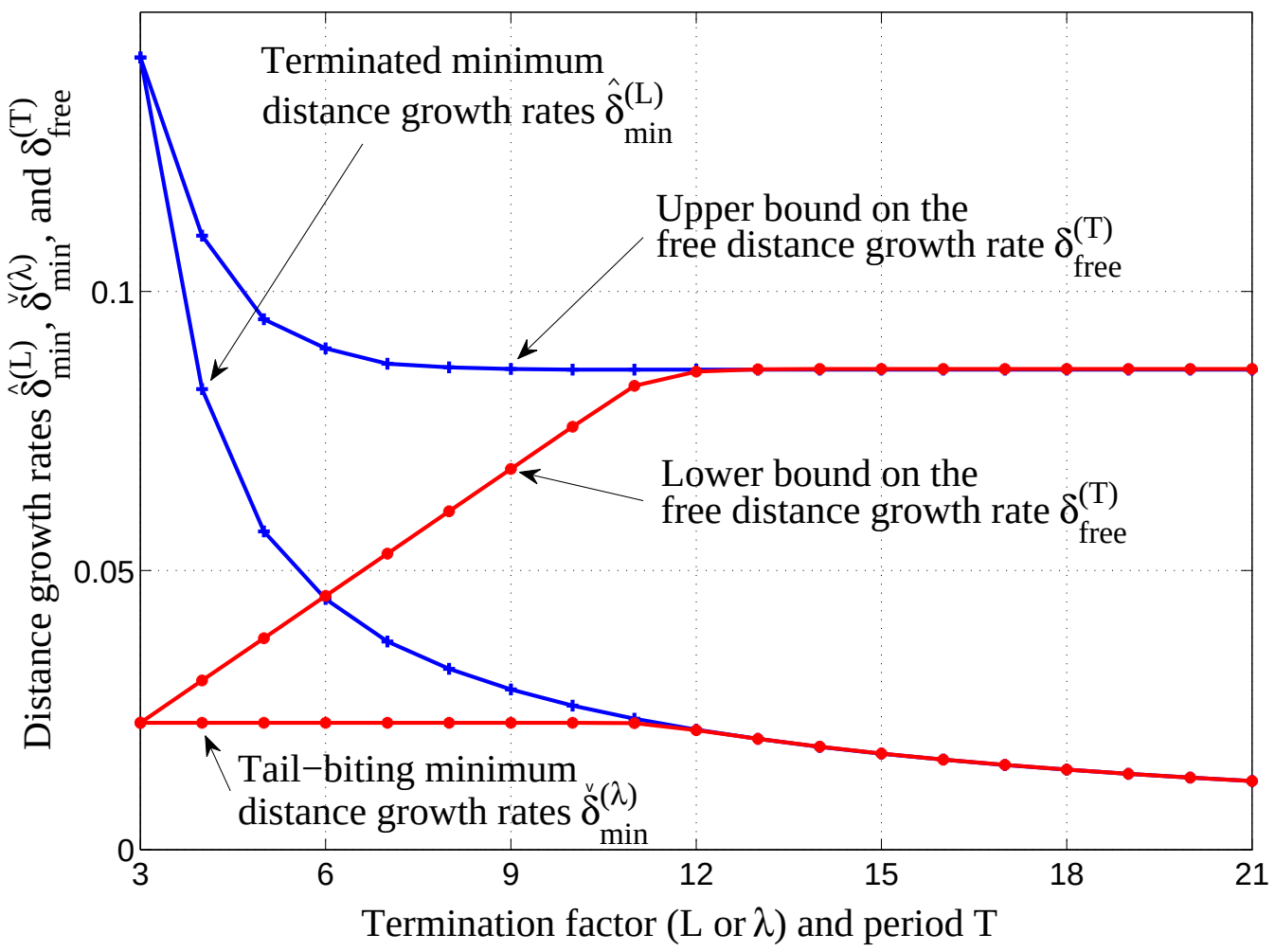

Fig. 1: Minimum distance growth rates of terminated and tail-biting LDPCC code ensembles with calculated upper and lower bounds on the free distance growth rate of the associated periodically time-varying LDPCC code ensembles.

Lower bounds on the free distance growth rates were calculated for a wide variety of $(J,K)$-regular and irregular LDPCC code ensembles in [17]. Using the technique detailed here, we can form upper bounds on the free distance growth rate that coincide numerically for sufficiently large $T$, giving us exact free distance growth rates. For example, we can bound the convolutional free distance growth rate of the $(4,8)$-regular ensemble as $0.1908 \leq \delta_{free}^{(T)} \leq 0.1908$ and the free distance growth rate for the rate $R = 2/3$, $(3,9)$-regular ensemble as $0.0186 \leq \delta_{free}^{(T)} \leq 0.0186$ for sufficiently large $T$ (again significantly larger than the corresponding block growth rates, see [16]). This general technique can be used to bound the free distance growth rate above and below for any regular or irregular periodically time-varying protograph-based LDPCC code ensemble.

## V. Trapping set analysis of LDPCC codes

In [7], MacKay and Postol discovered a "weakness" in the structure of the Margulis construction of a $(3,6)$-regular Gallager code. Described as *near-codewords*, these small graphical sub-structures existing in the Tanner graph of LDPC codes cause the iterative decoding algorithm to get trapped in error patterns. These weaknesses were shown to contribute significantly to the performance of the code in the error floor region of the BER curve. Richardson developed this concept in [8] and defined these structures as *trapping sets*.

[1] The free distance growth rate may also be calculated with respect to the encoding constraint length $\nu_e$, which corresponds to the maximum number of transmitted symbols that can be affected by a single nonzero block of information digits. For further details, see [16].

[2] The free distance growth rate $\delta_{free}^{(T)}$ that we bound from above using (5), by definition, is an existence-type lower bound on the free distance of most members of the ensemble, i.e., with high probability a randomly chosen code from the ensemble has minimum free distance at least as large as $\delta_{free}^{(T)} \nu_s$ as $\nu_s \to \infty$.

*Definition 1:* An $(a, b)$ general trapping set $\tau_{a,b}$ of a bipartite graph is a set of variable nodes of size $a$ which induce a subgraph with exactly $b$ odd-degree check nodes (and an arbitrary number of even-degree check nodes).

In order to calculate ensemble average general trapping set enumerators for protograph-based LDPC block code ensembles, we use the combinatorial arguments previously presented in [18]. The technique involves considering a two-part ensemble average weight enumerator for a modified protograph with the property that any $(a, b)$ trapping set in the original protograph is a codeword in the modified protograph.

### A. Trapping set growth rates

Let $\Delta = b/a = \beta/\alpha$, where $\alpha = a/n$, $\beta = b/n$, and $\Delta \in [0, \infty)$. As proposed in [18], we classify the trapping sets as $\tau_\Delta = \{\tau_{a,b} | b = \Delta a\}$. For each $\Delta$, we define $d_{ts}(\Delta)$ to be the $\Delta$-*trapping set number*, which is the size of the smallest, non-empty trapping set in $\tau_\Delta$. The two-part average ensemble average weight distribution can be used to test if the ensemble has the desirable property that the $\Delta$-trapping set number increases linearly with block length $n$ [18]. If this is the case, we can say that, with high probability, a randomly chosen code from the ensemble has a $\Delta$-trapping set number that is at least as large as $n\delta_{ts}(\Delta)$, where $\delta_{ts}(\Delta)$ is called the $\Delta$-*trapping set growth rate* of the ensemble. If this is true for all $\Delta \geq 0$, this implies that, for sufficiently large $n$, a typical member of the ensemble has no small trapping sets.

### B. Trapping set bounds for protograph-based LDPCC code ensembles

Consider once more the ensemble of periodically time-varying of LDPCC codes with rate $R = 1 - b_c/b_v$ and period $T$ derived from a convolutional base matrix $\mathbf{B}_{[0,\infty]}$ and the associated terminated LDPCC code ensemble with base matrix $\mathbf{B}_{[0,L-1]}$ and $L = T$.

*Theorem 2:* Consider a rate $R = 1 - b_c/b_v$ unterminated, periodically time-varying convolutional code ensemble with memory $m_s$, decoding constraint length $\nu_s = N(m_s + 1)b_v$, and period $T$ derived from $\mathbf{B}_{[0,\infty]}$. Let $\overline{d}_{ts}^{(L)}(\Delta)$ be the average $\Delta$-trapping set number of the terminated convolutional code ensemble with block length $n = LNb_v$ and termination factor $L$. Then the ensemble average $\Delta$-trapping set number $\overline{d}_{ccts}^{(T)}(\Delta)$ of the unterminated convolutional code is bounded above by $\overline{d}_{ts}^{(L)}(\Delta)$ for termination factor $L = T$ and any $\Delta \geq 0$, i.e.,

$$\overline{d}_{ccts}^{(T)}(\Delta) \leq \overline{d}_{ts}^{(T)}(\Delta) \quad \forall \Delta \geq 0. \tag{7}$$

*Sketch of proof.* The proof is a straightforward generalisation of the proof of Theorem 1. We first show that, for any periodically time-varying LDPCC code and associated terminated LDPCC code with termination factor $L = T$, and any $\Delta \geq 0$, any $(a, \Delta a)$ general trapping set in the terminated code is also an $(a, \Delta a)$ general trapping set in the convolutional code, i.e., the $\Delta$-trapping set number of the convolutional code $d_{ccts}^{(T)}(\Delta)$ is bounded above by the $\Delta$-trapping set number of the terminated code $d_{ts}^{(L)}(\Delta)$ for $L = T$ and any $\Delta \geq 0$. This can be shown by considering a pair of modified code ensembles where each check node is connected once to a distinct *auxiliary* variable node (see [18]). Crucially, there is a bijective mapping from the set of all $(a, b)$-general trapping sets in the original code to the set of all codewords in the modified code, and we can use a minimum distance-type argument to prove the result for the modified code. The ensemble average result $\overline{d}_{ccts}^{(T)}(\Delta) \leq \overline{d}_{ts}^{(T)}(\Delta)$ for all $\Delta \geq 0$ then follows directly. □

Again, we will henceforth drop the overline notation when discussing ensemble average $\Delta$-trapping set enumerators. Using (7) and a similar sequence of arguments to those presented in Section IV-B, we can form an upper bound on the $\Delta$-trapping set growth rate $\delta_{ccts}^{(T)}(\Delta)$ of the periodically time-varying LDPCC code ensemble as

$$\delta_{ccts}^{(T)}(\Delta) = \frac{d_{ccts}^{(T)}(\Delta)}{\nu_s} \leq \frac{\hat{\delta}_{ts}^{(T)}(\Delta)T}{(m_s + 1)}, \tag{8}$$

where $\hat{\delta}_{ts}^{(T)}(\Delta)$ is the $\Delta$-trapping set growth rate of the terminated LDPCC code ensemble with termination factor $L = T$ and base matrix $\mathbf{B}_{[0,T-1]}$ for any $\Delta \geq 0$. Similarly, a lower bound on $\delta_{ccts}^{(T)}(\Delta)$ was calculated in [19] using tail-biting LDPCC code ensembles as

$$\delta_{ccts}^{(T)}(\Delta) \geq \frac{\check{\delta}_{ts}^{(T)}(\Delta)T}{(m_s + 1)}, \tag{9}$$

where $\check{\delta}_{ts}^{(T)}(\Delta)$ is the $\Delta$-trapping set growth rate of the tail-biting LDPCC code ensemble with termination factor $\lambda = T$ and base matrix $\mathbf{B}_{tb}^{(T)}$ for any $\Delta \geq 0$.

### C. Numerical results

We continue our analysis of the $(3, 6)$-regular LDPCC code ensemble described in Section II-A. Since the unterminated convolutional code has rate $R = 1/2$, we calculate the upper bound on the $\Delta$-trapping set growth rate of the periodically time-varying LDPCC code ensemble as $\delta_{ccts}^{(T)}(\Delta) \leq \hat{\delta}_{ts}^{(T)}(\Delta)T/(m_s + 1)$ using (8) for termination factors $L = T \geq 3$. For $\Delta = 0, 0.01, 0.05$, Figure 2 displays the $\Delta$-trapping set growth rates $\hat{\delta}_{ts}^{(T)}(\Delta)$ of the terminated ensembles defined by $\mathbf{B}_{[0,L-1]}$ for $L = 3, \ldots, 18$ (calculated using techniques from [18]) and the associated upper bounds on the convolutional $\Delta$-trapping set growth rate $\delta_{ccts}^{(T)}(\Delta)$ for $L = T$. Also shown are the $\Delta$-trapping set growth rates $\check{\delta}_{ts}^{(T)}(\Delta)$ of the tail-biting ensembles defined by $\mathbf{B}_{tb}^{(\lambda)}$ for $\lambda = 3, 6, \ldots, 18$ and the associated lower bounds on the convolutional growth rates $\delta_{ccts}^{(T)}(\Delta)$ calculated using (9) that were obtained in [19].

Note that setting $\Delta = \beta/\alpha = 0$ corresponds to the minimum distance growth rate problem discussed in Section IV, and as a result, the curves corresponding to $\Delta = 0$ match those displayed in Figure 1. For $\Delta = 0.01$ and $\Delta = 0.05$ we observe the same behaviour: the $\Delta$-trapping set growth rates of the LDPC block code ensembles defined by $\mathbf{B}_{[0,T-1]}$ and $\mathbf{B}_{tb}^{(T)}$ are positive and decrease monotonically to zero as the termination factors tend to infinity. For each $\Delta$, the corresponding upper and lower bounds calculated for $\delta_{ccts}^{(T)}(\Delta)$ using (8) and (9) (respectively) coincide for $T \geq 12$ and decrease as $\Delta$ increases. The empirical data suggests that the bounds will remain equal and constant for $T > 18$.

As $\Delta$ ranges from 0 to $\infty$, the points $(\delta_{ts}(\Delta), \Delta\delta_{ts}(\Delta))$ trace out the so-called *zero-contour curve* for a protograph-based block code ensemble [18]. The zero-contour curves for the $(3,6)$-regular LDPC block code ensemble and the periodically time-varing LDPCC code ensemble with $T = 12$ are shown in Figure 3.[3] The Δ-trapping set growth rates are highlighted for $\Delta = 0.01$.

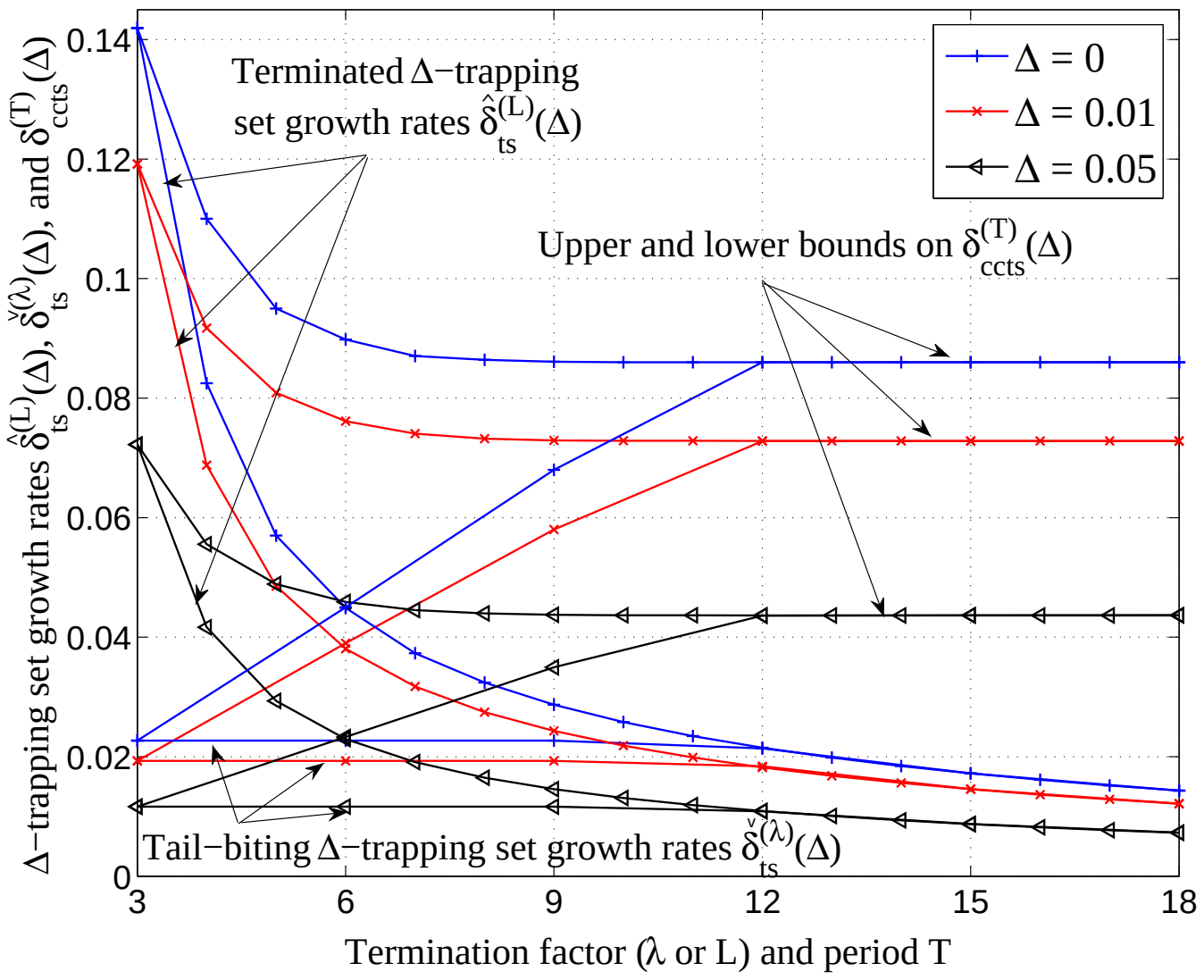

Fig. 2: Δ-trapping set growth rates of terminated and tail-biting LDPCC code ensembles with calculated upper and lower bounds on the Δ-trapping set growth rate of the associated periodically time-varying LDPCC code ensembles.

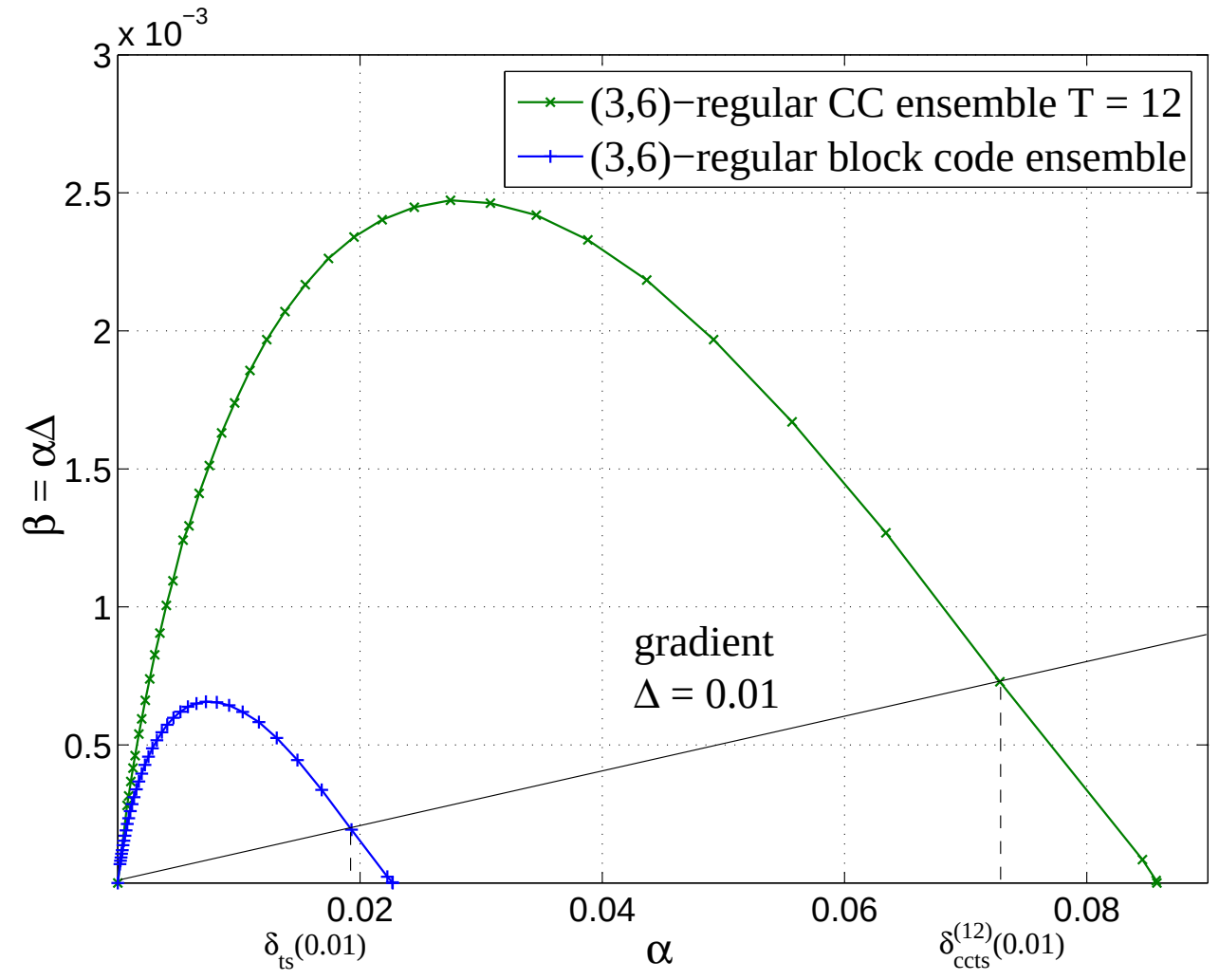

Fig. 3: Zero-contour curves of the $(3,6)$-regular LDPC block code ensemble and the $(3,6)$-regular LDPCC code ensemble.

For all $\Delta \geq 0$, $\delta_{ccts}^{(12)}(\Delta) > 0$, indicating that, for each class of $(a, b)$ general trapping set, the size of the smallest non-empty trapping set typical of most members of the ensemble is growing linearly with constraint length. Code ensembles with large Δ-trapping set numbers $d_{ccts}^{(T)}(\Delta)$ are the most interesting, since small trapping sets dominate iterative decoding performance in the error floor [8]. Thus we want the Δ-trapping set growth rate $\delta_{ccts}^{(12)}(\Delta)$ to exist and to be as large as possible, thus guaranteeing good iterative decoding performance in the error floor. Finally, we note that the convolutional growth rate $\delta_{ccts}^{(12)}(\Delta)$ exceeds the associated block growth rate $\delta_{ts}(\Delta)$ for all $\Delta \geq 0$.

## VI. Conclusions

In this paper we showed, using a protograph-based analysis of terminated LDPCC codes, that we can obtain an upper bound on the free distance growth rate of an ensemble of periodically time-varying LDPCC codes. We found that the bounds we obtain coincide with lower bounds previously obtained by analysing the minimum distance of ensembles of tail-biting LDPCC codes. This approach was then extended to obtain upper and lower bounds on the Δ-trapping set growth rates of ensembles of periodically time-varying LDPCC codes. Further, it was shown that the distance and Δ-trapping set growth rates of the LDPCC code ensembles exceed the growth rates of the corresponding LDPC block code ensembles on which they are based. The large minimum distance and trapping set growth rates obtained suggest that LDPCC codes will exhibit good iterative decoding performance in the error floor.

[3]For $T = 12$, the upper and lower bounds coincide for all calculated values of Δ. This enables us to plot an exact zero-contour curve, in contrast to the lower-bound zero-contour curve reported in [19].

## References

[1] A. J. Felström and K. Sh. Zigangirov, "Time-varying periodic convolutional codes with low-density parity-check matrices," *IEEE Trans. Inf. Theory*, vol. 45, no. 6, pp. 2181–2191, Sept. 1999.

[2] A. E. Pusane, R. Smarandache, P. O. Vontobel, and D. J. Costello, Jr., "Deriving good LDPC convolutional codes from LDPC block codes," *IEEE Trans. Inf. Theory*, vol. 57, no. 2, pp. 835–857, Feb. 2011.

[3] M. Lentmaier, A. Sridharan, D. J. Costello, Jr., and K. Sh. Zigangirov, "Iterative decoding threshold analysis for LDPC convolutional codes," *IEEE Trans. Inf. Theory*, vol. 56, no. 10, pp. 5274–5289, Oct. 2010.

[4] M. Lentmaier, G. P. Fettweis, K. Sh. Zigangirov, and D. J. Costello, Jr., "Approaching capacity with asymptotically regular LDPC codes," in *Proc. Inf. Theory and App. Workshop*, San Diego, CA, Feb. 2009.

[5] S. Kudekar, T. Richardson, and R. Urbanke, "Threshold saturation via spatial coupling: why convolutional LDPC ensembles perform so well over the BEC," in *Proc. IEEE Int. Symp. on Inf. Theory*, June 2010.

[6] A. Sridharan, D. Truhachev, M. Lentmaier, D. J. Costello, Jr., and K. Sh. Zigangirov, "Distance bounds for an ensemble of LDPC convolutional codes," *IEEE Trans. Inf. Theory*, vol. 53, no. 12, pp. 4537–4555, 2007.

[7] D. J. C. MacKay and M. S. Postol, "Weaknesses of Margulis and Ramanujan-Margulis low-density parity-check codes," *Electronic Notes in Theoretical Computer Science*, vol. 74, pp. 97–104, 2003.

[8] T. J. Richardson, "Error-floors of LDPC codes," in *Proc. Forty-first Annual Allerton Conference*, Allerton, IL, Sept. 2003.

[9] D. G. M. Mitchell, A. E. Pusane, K. Sh. Zigangirov, and D. J. Costello, Jr., "Asymptotically good LDPC convolutional codes based on protographs," in *Proc. IEEE Int. Symp. on Inf. Theory*, July 2008.

[10] D. Truhachev, K. Sh. Zigangirov, and D. J. Costello, Jr., "Distance bounds for periodically time-varying and tail-biting LDPC convolutional codes," *IEEE Trans. Inf. Theory*, vol. 56, no. 9, pp. 4301–4308, 2010.

[11] J. Thorpe, "Low-density parity-check (LDPC) codes constructed from protographs," Jet Prop. Lab., INP Progress Report 42-154, Aug. 2003.

[12] W. S. Massey, *Algebraic Topology: an Introduction*. New York: Springer-Verlag, Graduate Texts in Mathematics, Vol. 56, 1977.

[13] G. Solomon and H. C. A. Tilborg, "A connection between block and convolutional codes," *SIAM Journal on Applied Mathematics*, vol. 37, no. 2, pp. 358–369, Oct. 1979.

[14] H. H. Ma and J. K. Wolf, "On tail biting convolutional codes," *IEEE Transactions on Communications*, vol. 34, no. 2, pp. 104–111, Feb. 1986.

[15] D. Divsalar, "Ensemble weight enumerators for protograph LDPC codes," in *Proc. IEEE Int. Symp. Inf. Theory*, Seattle, WA, July 2006.

[16] D. G. M. Mitchell, A. E. Pusane, and D. J. Costello, Jr., "Minimum Distance and Trapping Set Analysis of Protograph-based LDPC Convolutional Codes," in preparation, 2011.

[17] D. G. M. Mitchell, A. E. Pusane, N. Goertz, and D. J. Costello, Jr., "Free distance bounds for protograph-based regular LDPC convolutional codes," in *Proc. Int. Symp. on Turbo Codes and Rel. Topics*, Aug. 2008.

[18] S. Abu-Surra, W. E. Ryan, and D. Divsalar, "Ensemble trapping set enumerators for protograph-based LDPC codes," in *Proc. Forty-fifth Annual Allerton Conference*, Allerton, IL, Sept. 2007.

[19] D. G. M. Mitchell, A. E. Pusane, and D. J. Costello, Jr., "Trapping set analysis of protograph-based LDPC convolutional codes," in *Proc. IEEE Int. Symp. Inf. Theory*, Seoul, Korea, July 2009.